# Managing & Analyzing Large Volumes of Dynamic & Diverse Data

Okal Christopher Otieno
Department of Information Technology
Mount Kenya University
Nairobi, Kenya

**Abstract:** This study reviews the topic of big data management in the 21st-century. There are various developments that have facilitated the extensive use of that form of data in different organizations. The most prominent beneficiaries are internet businesses and big companies that used vast volumes of data even before the computational era. The research looks at the definitions of big data and the factors that influence its access and use for different persons around the globe. Most people consider the internet as the most significant source of this data and more specifically on cloud computing and social networking platforms. It requires sufficient and adequate management procedures to achieve the efficient use of the big data. The study revisits some of the conventional methods that companies use to attain this. There are different challenges such as cost and security that limit the use of big data. Despite these problems, there are various benefits that everyone can exploit by implementing it, and they are the focus for most enterprises.

## 1. INTRODUCTION

The growth of businesses around the globe has always motivated the owners and managers to try and ensure their enterprises keep up with the technological trends that are equally developing at a fast pace. This progress is especially important in the field of business analytics according to past research. These studies reveal that 97% of the companies whose revenues exceed $100 million are embracing technology into their practices of analysis on their operations [3].

These developments have provided the background for using information technology (IT) in organizations around the world. Computational techniques are necessary for every sector as they provide an avenue that quickens the operations as well as ensuring the efficiency and precision of the outcomes [10]. This action is equally significant in the management of data structures in the organizations and any other operations that involve the sourcing, analysis and use of information [14].

These changes in business trends open up to the topic of this research that is, the management and analysis of big data. These practices involve the management of the strategies that companies use to get the data from the fields and create databases where they can manage it [3]. They also define the methodologies that the companies will use to achieve the best outcomes from their data.

In this research paper, the investigation focuses on the definition of the contexts that describe big data systems. It will also outline the importance of these systems in business operations and indulge into their management. The aim is to determine the processes that define the management and analysis of large volumes of data that is diverse and dynamic in nature. The study will provide an overview of the challenges that businesses go through to attain the efficiency that is their target in using this type of structures.

## 2. LITERATURE REVIEW

### 2.1 What is Big Data?

Big Data refers to data sets of vast volumes that can provide for computational manipulations and analysis to reveal the trends and associations between various variables of organizational operations [22]. There is an association between these forms of data with the Internet and cloud computing platforms especially proposing it as a 21st-century technology. This relationship goes to these avenues providing the ground for the use and growth of big data systems. It occurs through the online platforms and small internet startups that have practiced it in very extensive use [6]. This connection originates from the ability of a business to operate a standalone infrastructure that big data provides. Some examples of companies that use these types of information in their systems are internet search engines and social networking websites [19].

Big data is the information that organizations use to determine most aspects of consumer and stakeholder behavior [8]. The size that defines the context of big data keeps growing with the advances in the infrastructure and the technologies that people use to handle it [8]. In the past people used mainframe computers whose storage capacities limited their capabilities and, therefore, their definition for big data would be just a small fraction of the current systems. In some instances, startups and the Internet platforms can see this as an innovation but larger companies have been doing big data from their inception (Davenport & Dyche, 2013). These enterprises have been handling large volumes of information in traditional ways that technology has improved tremendously.

The management and analysis of big data rely on the IT infrastructure to exploit the information that unstructured data contains and it is the form that dominates the globe [23]. To counter on the recent definitions of big data, some point that there is no sense of the distinction between any other forms. On the description of its size, some argue that data will continue growing and, therefore, one cannot distinguish big data in terms of size [7]. Therefore, the exact limits of the size that defines the term 'Big Data' lack any clarity as the amounts will keep changing with the advancements in technology. According to some research, one considers any information as large if they cannot handle it using the tools and software that they count as of ordinary use [7]. The infrastructural inadequacy to handle this form of data arises from the complexity that it contains and requires better methodologies to handle the larger scales that it presents.

2.2 Big Data Mining

One of the primary attributes of big data is that its sources have independent events that no one can monitor at any particular instances [24]. This nature presents different challenges that necessitate complicated methods of sourcing the information in the fields into the data sets. The name given to this exercise is data mining and in summary, it is the process of extracting pertinent information from those big datasets that are autonomous [7]. The sets consist of high levels of variability, velocity and gigantic volumes that are hard to predict due to the manner of growth they exhibit. The heterogeneous state of the variables is among the factors that increase this problem [17].

The growth of IT and the use of the internet have always had an excellent association with big data in the current society (Davenport & Dyche, 2013). This relationship has established cloud computing as one of the areas that organizations source their big data [12]. The development of the technologies in this sector has enabled people and organization to interconnect and share information on various platforms. Social media is among the top enablers of big data mining strategies for different groups especially information on human behavior [17]. Improvements in parallel programming are also an important development that facilitates various techniques of sourcing large data sets [18]. Some of the tools that are in common use for this purpose are software applications like Hadoop and MapReduce. They are in a position to handle the exponential growth of data in the globe [11].

2.3 Management and Analysis of Large Volumes of Data

There are different benefits that the access to big data has brought to various enterprises including the elimination of guesswork in decision-making [20]. The growth that data structures are growing around the world has enabled the possibility of accessing any information through different platforms. This expansion in the amounts of information people can access requires them to establish a system that allows proper monitoring and analyzing the data and this demand keeps growing [21]. Therefore, large-volume data management involves the processes of dealing with both structured and unstructured datasets. They include properly organizing, governing and administering vast volumes of such data sets for efficient utilization [4]. Most enterprises use database management systems (DBMS) to outline this function in their operations [1].

*2.3.1 Best Practices in Big Data Analysis and Management:* It requires proper management techniques to ensure that an organization can benefit from its adoption of big data strategies, especially in the computing platforms. These practices concern all the aspects of an enterprise's operations revolving around their culture and implementation methodologies. They also ensure that the organization can cope with the different changes that the environment of activity can send its way from period to another.

Establishing governing standards for the usability of the big data infrastructure is one of the practices to ensure efficiency in the organization [16]. The managers should ensure that they establish stringent policies that can enable the people in the business to observe a particular order in using the databases. The business should have policies that govern the access that every level of employee in the organization has to the data sets. The scope of these guidelines should begin from the basic oversight exercises right up to the overall monitoring processes of information use [16]. The aim of outlining these policies is to ensure that the people in the organization are only able to make the changes that the managers authorize. Therefore, it lowers the probability that people can just tamper with the data as it is likely to increase the variability that makes it harder to analyze. It is also one of the ways that they can make sure their database is secure from manipulation. It is also important to establish the difference between these policies and those of other operations in the enterprise [22].

Planning for quality in a big data system is also another important practice in analyzing and managing large volumes of information [13]. The use of big data in an organization has the potential to mislead decision-makers. It results from the lack of adequate opportunities to perform quality checks and simultaneously ensuring real-time access to the events that are occurring in the relevant perspectives [13]. The best management practices should aim to optimize between the two areas. It takes precious time to check data for credibility and reliability and this may hinder the operations of acquiring more from the field. It is wiser, therefore, for the enterprise to establish a link between the two activities. They should ensure that they perform tests and corrections on the data at the fastest possible rates of processing to enable them to continue with the activities of mining [13]. The organization requires the establishment of methods that can allow them to cleanse the information on the go, for example, using information filters and sorting techniques that work alongside data mining [13].

Agility to the changing trends in the data platforms is an important consideration in managing its larger volumes [13]. The changes are too rapid and require that the analysts be keen on the shifts; therefore, there is no sufficient time to build a long-lasting strategy to handle the information [13]. The users have to stay alert and focus on obtaining newer facts and analytical methods instead of focusing their energy on the current structures [22]. The implementation of management decisions should focus on the foreseeable future as the patterns are shifting rapidly and would be difficult to predict their long-term outcomes [13]. It also requires that the people in charge are ready for any disruptions that are likely from various sources in their environments. The interruptions may arise from a change in the technology that they use, a shift in the variables of focus, or even the evolution of programming methodologies that are available [13]. These events demand high levels of flexibility to accommodate the shifts that the personnel may experience in outlining their mandates of big data management.

Efficiency and reliability the storage capabilities an enterprise has at its dispensation is also a very significant consideration in managing large volumes of data [13]. This segment also goes hand-in-hand with the requirement of infrastructural flexibility. It is good to ensure that one has the best facilities that can handle the volume of information that they are likely to access and keep in their systems. The users must establish the use of flexible technologies in the form of hardware and software to manage their information, both in accessing and storing it [13]. In this aspect, the organization should ensure they are operating in compliance with the legal frameworks of their territories. For example, they should have sufficient licensing for their activities and materials or opt for the open source equipment and software to reduce these costs. It is also important that they maintain a good schedule on backing up their data and archiving other valuable information in their operations [13]. Security is another concern that is pertinent to this practice, but the solution can be in ensuring proper policy implementation and constant monitoring of the activities.

Data modeling is another important aspect of big data management for different users [13]. It refers to the activities that are likely to influence the logical and physical characteristics of the information in the enterprise [13]. The best proposal on data modeling for large-volume operations is to divide it into dimensions and deal with it in those subdivisions. Matching the information to the real world situation can be a challenge if one does not consider proper models for their data. This process is also important in ensuring that the analysis procedures have a sufficient facilitation and that the results have a higher precision capacity [13]. The dimensioning strategy is also important in ensuring good governance of big data, especially in handling the policies of access to different segments of a database.

*2.4 Challenges in Big Data Management*

The first problem in the analysis and management of big data derives from the nature of the information it contains [9]. The datasets are in vast scales that include heterogeneous rules and patterns whose characteristics exhibit high variability [9]. These issues present a challenge right from the exercises of data mining up to its distribution and usage in the relevant areas. Sometimes it will require sophisticated algorithms and programming procedures to enable the organizations to use this form of data efficiently. The amount of variation in the source variables makes it difficult for managers and developers to make any predictions about the data they have due to its enormous scale as well [9]. The algorithms are especially important in establishing searches for particular information. If the procedures are not adequate in terms of speed and accuracy, they will slow down the processes and reduce the precision of the data that the organization has [9]. Developing these methods for use in an organization is time-consuming and also very costly and in turn poses a constraint for implementation in smaller scales of business [9].

The amounts of variability pose another challenge of the increase of workload in the organization. The heterogeneity of the variables from the sources of big data implies that an organization cannot use a single database structure to monitor and analyze the activities [25]. Therefore, it has to administer more labor into creating a newer infrastructure to adjust to the changes in patterns of the raw data. This trend also calls for a change in analytical methods every time they experience a particular shift in the sources. These events will increase the amount of work that the workers have to handle and also its diversity. It then requires the enterprise to increase its investment into these processes, and that equally raises the cost of operation [5].

Security is another challenge that is very frequent in the arena of implementation and management of large volumes of data [2]. The information that most big data infrastructures contain is highly sensitive and can motivate cyber attacks from different people to gain access to it. Most of the information is very critical for organization processes such as marketing and finances that people can develop an interest in to complete malicious events [15] The diversity of sources that companies get their data also brings the challenge of its legitimate use and the protection of privacy for its customers [15] The access to an enterprise's information in such situation can cause havoc as it usually has the personal data about employees and customers. The process of protecting the databases from cyber attacks that can lead to dangerous manipulations comes with many implications. It leads to an increase in the managerial workload and costs of operation for the business in question.

Cost is another challenge that overlaps through all the above limitations to big data. The cost of implementing the infrastructure necessary to handle these operations is high. In each of the activities and the various challenges that come up, the organization will always need to expend funds to deal with the problems. Though, the advances in technology and social applications of large data, there is a promise that all this will become affordable at a particular period [20]. In the meantime, such technologies as cloud computing can fill in the gap by lowering the infrastructural expenses that most organizations have to incur.

### 3. CONCLUSION

There are different benefits that people can get from implementing and using large volumes of data. It is beneficiary to both educational engagements and business decision-making for various environments of application around the globe. The only way to ensure that one exploits these benefits is if they practice the best methodologies for analysis and management of the information that is under their mandate. These actions are specifically significant for business operations as the data they have exhibits a lot about their target markets. It also has the potential of influencing consumer behavior and is wise if they can perform proper managerial techniques on it.

There are different challenges in the sourcing, implementation and usage of big data in the current environment. The largest problem is the cost of adoption for such data especially for people who have not established the cases for business use of the data. Other challenges are in the nature of the variables that the persons have to observe with the primary focus on the variability, velocity and volume of the information that is available. Rapid changes do not allow the users to establish a standard way of sourcing and measuring the information, and this is a big challenge especially on cost.

There is an advantage to the diversity of the sources of big data despite the challenges they pose. This diverse nature of the information can be a significant source of decision-making data in an organization primarily because it reveals the variations in consumer behavior. With proper analysis of the findings in the target markets, the enterprise can benefit from such information. There is the challenge of ensuring that one is up to date about the different events about their sources and the rapid shift in their nature. It brings the difficulty of predictions that long-term decision-makers requires. Therefore, it is through proper implementation of the best management practices that an entity can benefit from such events to make decisions for the foreseeable future.

*References*
[1]. Borkar, Vinayak, Michael J. Carey, and Chen Li. "Inside Big Data management: ogres, onions, or parfaits?." *Proceedings of the 15th International Conference on Extending Database Technology*. ACM, 2012.
[2]. Cardenas, Alvaro A., Pratyusa K. Manadhata, and Sreeranga P. Rajan. "Big Data Analytics for Security." *IEEE Security & Privacy* 6, 2013, 74-76.
[3]. Chen, Hsinchun, Roger HL Chiang, and Veda C. Storey. "Business Intelligence and Analytics: From Big Data to Big Impact." *MIS quarterly* 36.4, 2012, 1165-1188.
[4]. Chen, Jinchuan, et al. "Big data challenge: a data management perspective." *Frontiers of Computer Science* 7.2, 2013, 157-164.
[5]. Chen, Yanpei, Sara Alspaugh, and Randy Katz. "Interactive Analytical Processing in Big Data Systems: A Cross-Industry Study of Mapreduce Workloads." *Proceedings of the VLDB Endowment* 5.12, 2012, 1802-1813.
[6]. Davenport, Thomas H., and Jill Dyché. "Big Data in Big Companies." May 2013.
[7]. Fan, Wei, and Albert Bifet. "Mining Big Data: Current Status, and Forecast to the Future." *ACM sIGKDD Explorations Newsletter* 14.2, 2013, 1-5.
[8]. Fisher, Danyel, et al. "Interactions with Big Data Analytics." *Interactions* 19.3, 2012, 50-59.
[9]. Fujimaki, Ryohei, & Satoshi Morinaga. "The Most Advanced Data Mining of the Big Data Era." *NEC Technical Journal* 7.2, 2012, 91.
[10]. Ivan, Ion, Cristian Ciurea, and Sorin Pavel. "Very Large Data Volumes Analysis of Collaborative Systems with Finite Number of States." *Journal of Applied Quantitative Methods* 5.1, 2010, 14-28.
[11]. Jaseena, K. U., and Julie M. David. "Issues, Challenges, and Solutions: Big Data Mining." 2014, 132-140.

[12]. Kepes, Ben. "How to Utilize Cloud Computing, Big Data and Crowdsourcing for an Agile Enterprise." *Gigaom Research*, 2014, 1-17.
[13]. Kimball, Ralph. "Newly Emerging Best Practices for Big Data." *Kimball Group*, 2015, 1-13.
[14]. Knobbe, Arno, et al. "InfraWatch: Data Management of Large Systems for Monitoring Infrastructural Performance." *Advances in Intelligent Data Analysis IX*. Springer Berlin Heidelberg, 2010. 91-102.
[15]. Lafuente, Guillermo. "The Big Data Security Challenge." *Network Security* 2015.1 (2015): 12-14.
[16]. LaValle, Steve, et al. "Big Data, Analytics and the Path From Insights to Value." *MIT SLOAN Management Review* 21, 2013.
[17]. Lin, Jimmy, and Dmitriy Ryaboy. "Scaling big data mining infrastructure: the twitter experience." *ACM SIGKDD Explorations Newsletter* 14.2, 2013, 6-19.
[18]. Moens, Sandy, Emin Aksehirli, and Bart Goethals. "Frequent itemset mining for big data." *Big Data, 2013 IEEE International Conference on*. IEEE, 2013.
[19]. Pearson, Travis, and Rasmus Wegener. "Big Data: The Organizational Challenge." *Bain Co*, 2013.
[20]. Purdue University. Challenges and Opportunities with Big Data. *Purdue University*, 2015, 1-17.
[21]. Rabl, Tilmann, et al. "Solving Big Data Challenges for Enterprise Application Performance Management." *Proceedings of the VLDB Endowment* 5.12, 2012, 1724-1735.
[22]. Russom, Philip. "Managing big data." *TDWI Research. TDWI Best Practices Report* , 2013.
[23]. Shelley, Phil. "Big Data Spectrum." *Infosys*, 2012, 1-57.
[24]. Wu, Xindong, et al. "Data mining with Big Data." *Knowledge and Data Engineering, IEEE Transactions on* 26.1, 2014, 97-107.
[25]. Zhu, Yuqing, et al. "Bigop: Generating Comprehensive Big Data Workloads as a Benchmarking Framework." *Database Systems for Advanced Applications*. Springer International Publishing, 2014.